\documentclass[twoside,slac_one]{revtex4}
\usepackage{graphicx}
\usepackage{fancyhdr}
\usepackage{amsmath} 
\usepackage{bm}
\usepackage{amsxtra}
\usepackage{amssymb}
\usepackage{amsthm}
\usepackage{latexsym}
\usepackage{lscape}

\newcommand{\p}{I\!\!P}

\newcommand{\Lb}{\left(}
\newcommand{\Rb}{\right)}
\def\pom{{I\!\!P}}
\def\reg{{I\!\!R}}

\begin{document}

\title{A Model for Soft Interactions motivated by AdS/CFT and QCD}

\author{E. Gotsman}
\affiliation{School  of Physics and Astronomy, Tel Aviv University,
 Tel Aviv, Israel}
\footnote{work done in collaboration with E.M. Levin and U. Maor}
\begin{abstract}
Monte Carlo generators which were tuned for energies up to that of the
 Tevatron, are found wanting when extended to LHC energies. We construct
 a model that satisfies the theoretical requisites of high energy
 soft interactions, based on two conjectures: (i) the results of the Ads/CDF
correspondence for N = 4 SYM, and (ii) the requirement of matching with
high energy QCD. In keeping with these postulates, we assume that the
soft Pomeron intercept is relatively large, and the slope of the Pomeron
trajectory is equal to zero. We derive analytical formulae that sum both
 the enhanced and semi-enhanced diagrams for elastic and diffractive
amplitudes. Parameters of the model are obtained by fitting to experimental
data, up to and including the Tevatron energies, and we predict cross
 sections at all energies accessible at the LHC and beyond.
 Predictions of the model are in agreement with measured values obtained by
CMS, ATLAS and ALICE for $\sigma_{inel}$ and for $\sigma_{incl}$. We compare
our results with experimental data and competing models.

\end{abstract}    

\maketitle

\thispagestyle{fancy}


\section{Introduction}
 The complete  algebraic treatment and results of our model are 
contained in GLM \cite{GLM} and GLMI \cite{GLMI}. \\
The following were our guiding criteria for constructing the GLM model.
\begin{center}
\begin{itemize}
\item
The model should be built using Pomerons and Reggeons.
\item
 The intercept of the Pomeron should be relatively large.
 In AdS/CFT correspondence  we
expect 
$\Delta_{\p} = \alpha_{\p}(0) - 1$ =
1 - 2/$\sqrt{\lambda} \approx$ 0.11 to 0.33.  The estimate for
$\lambda $  from the cross section for  multiparticle production
 as well as from DIS at HERA  is $\lambda$ = 5 to 9.
\item
 $\alpha'_{\p}(0) \,=\,0$.
\item
A large Good-Walker component is expected, as in the AdS/CFT approach
 the main contribution to shadowing corrections comes from elastic
scattering and diffractive production.
\item
The Pomeron self-interaction should be small
 (of the order of $2/\sqrt{\lambda}$ in AdS/CFT correspondence), and much
smaller than
 the vertex of interaction of the Pomeron with a hadron, which is
 of the order of $\lambda$.
\item
The last requirement follows  from the natural
matching with perturbative QCD: where the only  vertex that
contributes is the triple Pomeron vertex.
\end{itemize}
\end{center}

\section{Fundamentals of the GLM model}

\subsection{The Good-Walker mechanism}
 In our approach diffractive dissociation is taken into account in the 
framework of a two channel model, which we developed in \cite{GW}.
This enables us to express all the relevant cross sections,
$\sigma_{tot}$,$\sigma_{el}$,$\sigma_{sd}$,$\sigma_{dd}$ and the forward 
elastic slope $B_{el}$ in terms of  three basic amplitudes $A_{11}$(s,b),
$A_{12}$(s,b), $A_{22}$(s,b) and a mixing parameter $\beta$.
Satisfying the unitarity constraints we can parametrize the scattering 
amplitudes in the following
simple form:
 \begin{equation}
A_{i,k}(s,b)=i \Lb 1 -\exp\Lb - \frac{\Omega_{i,k}(s,b)}{2}\Rb\Rb,
\label{eq-ampl}
\end{equation} 
our parametrization of the opacities $\Omega_{i,k}(s,b)$ will be discussed 
later. A deficiency of the Good-Walker approach is that it only accounts 
for diffraction in the region of small mass. 

\subsection{Enhanced and semi-enhanced diagrams}
To calculate the contribution of produced large mass in diffraction 
dissociation, it is necessary to sum diagrams which contain 
Pomeron-Pomeron interactions. Examples of these diagrams are shown in 
Fig.1 
\begin{figure}[ht]
\includegraphics[width=80mm]{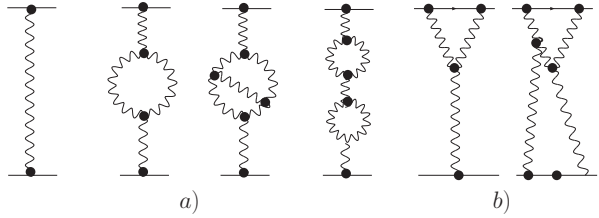}
\caption{Example of enhanced and semi-enhanced diagrams: a) examples of 
enhanced diagrams which  contribute to renormalizition of the Pomeron 
propagator; b) examples of semi-enhanced diagrams which contribute to the 
renormalization of the $\p$-p vertex.} \label{figure_1}
\end{figure}
\par To sum  the enhanced and semi-enhanced diagrams, we employ the MPSI 
approximation (where only large $\p$ loops of rapidity size O(Y) 
contribute). This allows us to obtain a closed expression for the Pomeron 
Green's function
\begin{equation}
G_{\p} \left(Y\right)\,\,\,=\,\,\,1\,\,-\,\,\exp \Lb \frac{1}{T(Y)}\Rb\,
\frac{1}{T(Y)}\,\,
\Gamma\Lb 0,\frac{1}{T(Y)} \Rb,
\label{eq-pgf}
\end{equation}
   where $\Gamma(0,x)$ is the incomplete gamma function,
 $T(Y) = \gamma \exp\Lb\Delta_{\p}Y\Rb$ and $\gamma$ denotes the amplitude
of the parton (dipole) interaction with the target at arbitrary rapidity 
Y.

 The summed amplitude can then be expressed in terms of $G_{\p}(T(Y))$
\begin{equation}
A_{i,k}\Lb Y; b\Rb\,\,\,= \,\, 1 \,\,\,- \,\,\exp\left\{ - \,\,\frac{1}{2}  
\,  
\int
d^2 b' \, \nonumber
\,\,\, \frac{\Lb \tilde{g}_i\Lb\vec{b}'\Rb\,\tilde{g}_k\Lb\vec{b} -  
 \vec{b}'\Rb\,G_{\p}(Y)\Rb}{ 
1\,+\,G_{\p}(Y)\,\left[\tilde{g}_i\Lb\vec{b}'\Rb +
 \tilde{g}_k\Lb\vec{b} - \vec{b}'\Rb\right]}\right\}
\label{eq-amgy}
\end{equation}   
where
\begin{equation}
\tilde{g}_{i}(\vec{b}) = g_{i}(\vec{b}) /\sqrt{\gamma}
\label{eq-tilde}
\end{equation}
\section{Predictions of our model}
  To determine  values for the parameters in our model we made 
a least squares fit,  over the ISR -Tevatron energy range, 
 to 55 data points . We obtain a $\chi^{2}$/d.o.f. = 0.86. The 
values of the 
parameters are given in \cite{GLM}.

The results for our total cross section and elastic cross section are shown in 
Fig.2, other cross sections appear in 
Fig.3.

\begin{figure*}[ht]
\centering
\begin{tabular}{c c c}
\includegraphics[width =70mm]{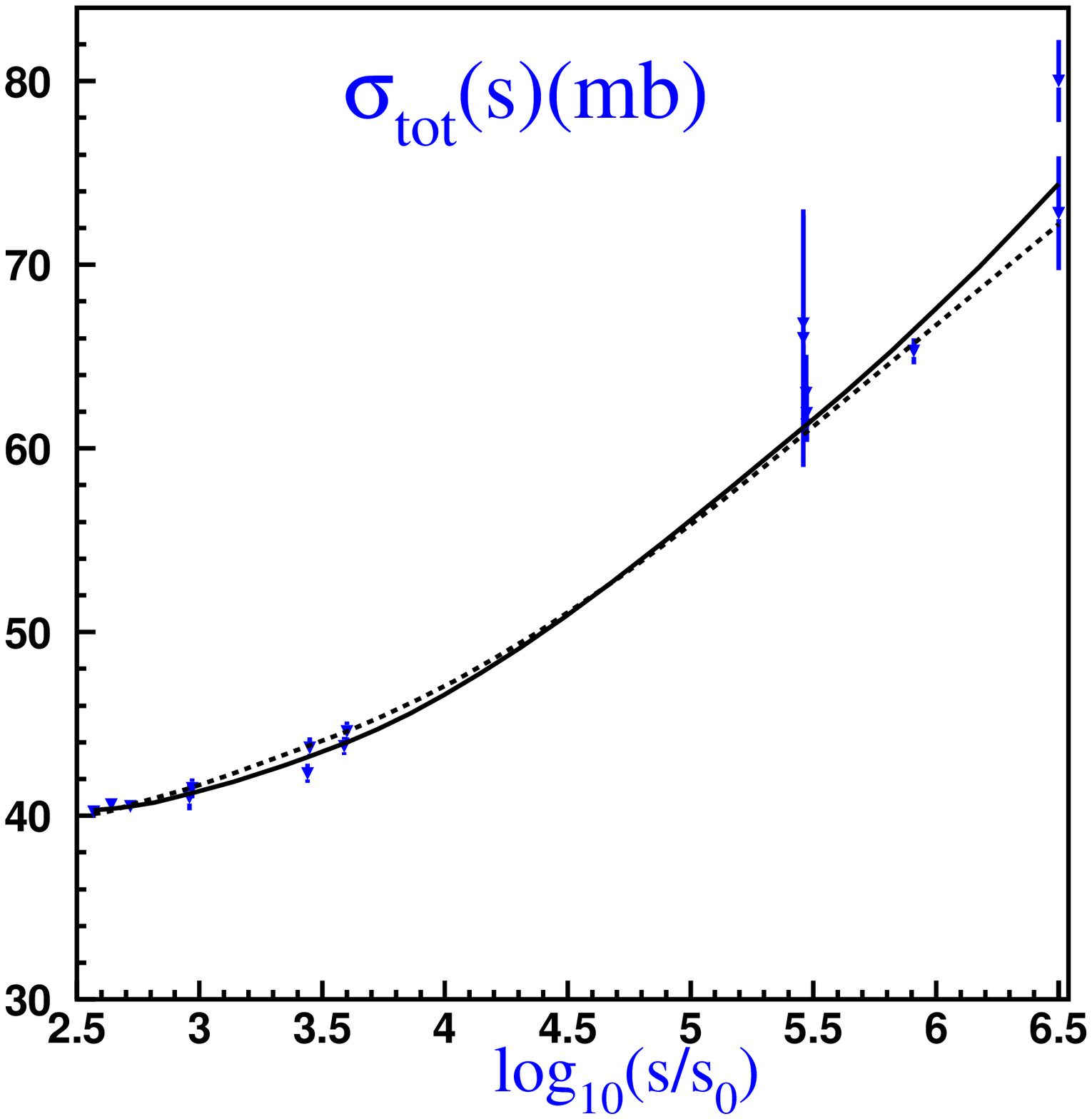} &\,\,\,\,&
\includegraphics[width =70mm]{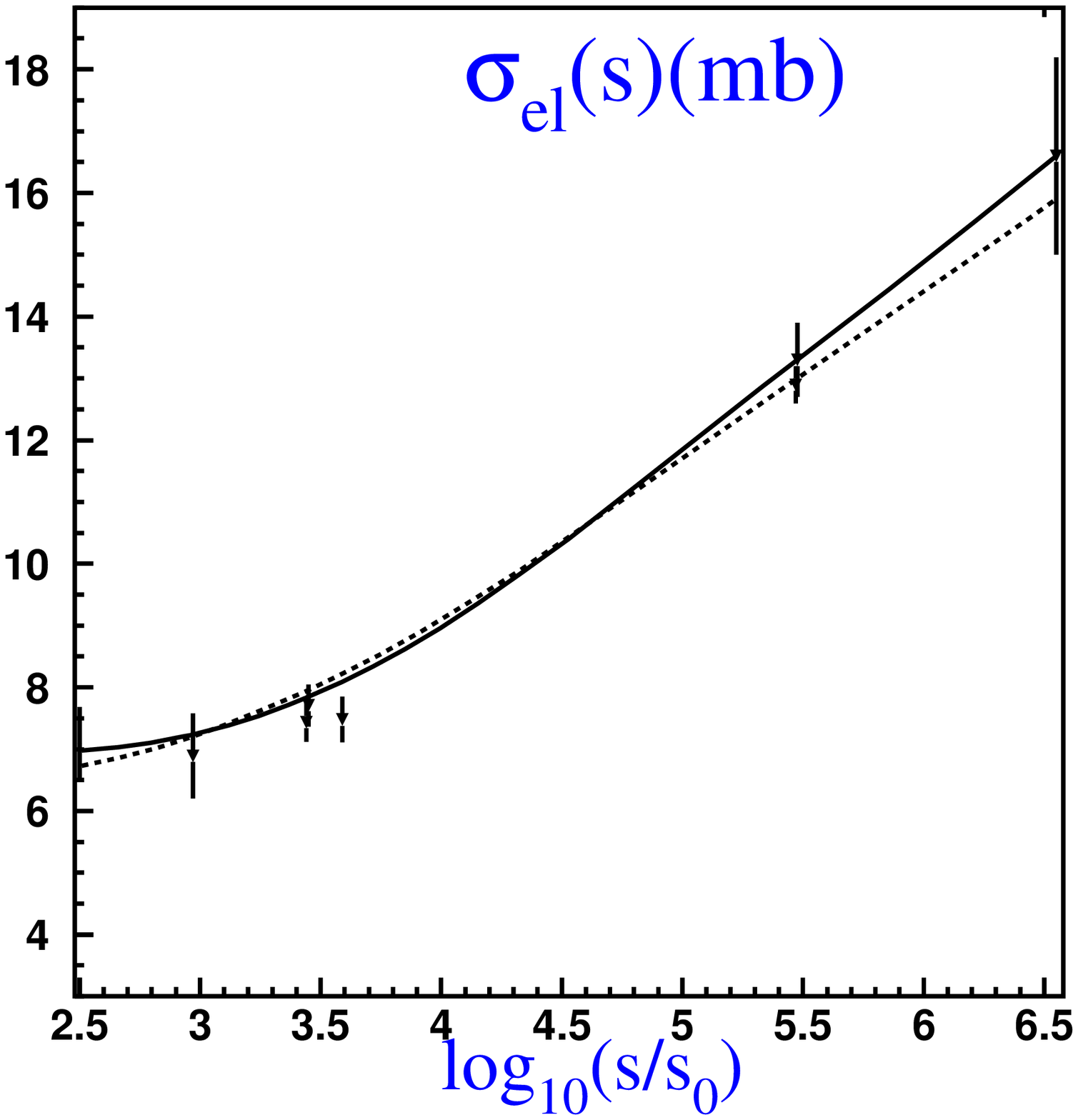}
\end{tabular}
\caption{GLM results for total cross  and elastic cross sections}
\label{crost_fig}
\end{figure*}

\subsection{Comparison with other models on the market}
Kaidalov and Poghosyan \cite{Kai} attempt to describe the data on soft 
diffraction by taking into account all possible non-enhanced absorptive 
corrections to three Pomeron and Reggeon vertices and loop diagrams. They 
apply AGK rules for calculating the discontinuity of the matrix element, 
and the generation of the optical theorem for the case of multi-Pomeron 
exchanges. They adopt the following parametrization for the Pomeron 
trajectory $\alpha_{\p}$ = 1.17 + 0.25t, and for the f and $\omega$ 
Reggeons, $\alpha_{f}$ =0.7 + 0.8t and $\alpha_{\omega}$= 0.4 + 0.9t. 

\par Ostapchenko \cite{Ostap} has made a comprehensive calculation in the 
framework of Reggeon field theory, based on the resummation of both 
enhanced and semi-enhanced Pomeron diagrams. To fit the total and 
diffractive cross sections he assumes two Pomerons: i) a "soft" Pomeron,
$\alpha^{soft}$ = 1.14 + 0.14t and ii) a "hard" Pomeron,
$\alpha^{hard}$ =1.31 + 0.085t.

\par The Durham group \cite{KMR} have a model which is similar in spirit 
to GLM, the main difference lies in their technique of summing the 
"Pomeron loop" diagrams, and their recent inclusion of $k_{t}$ evolution.

\par The  results of the four models are compared in Fig.3.

\begin{figure*}[ht]
\centering
\includegraphics[width=135mm]{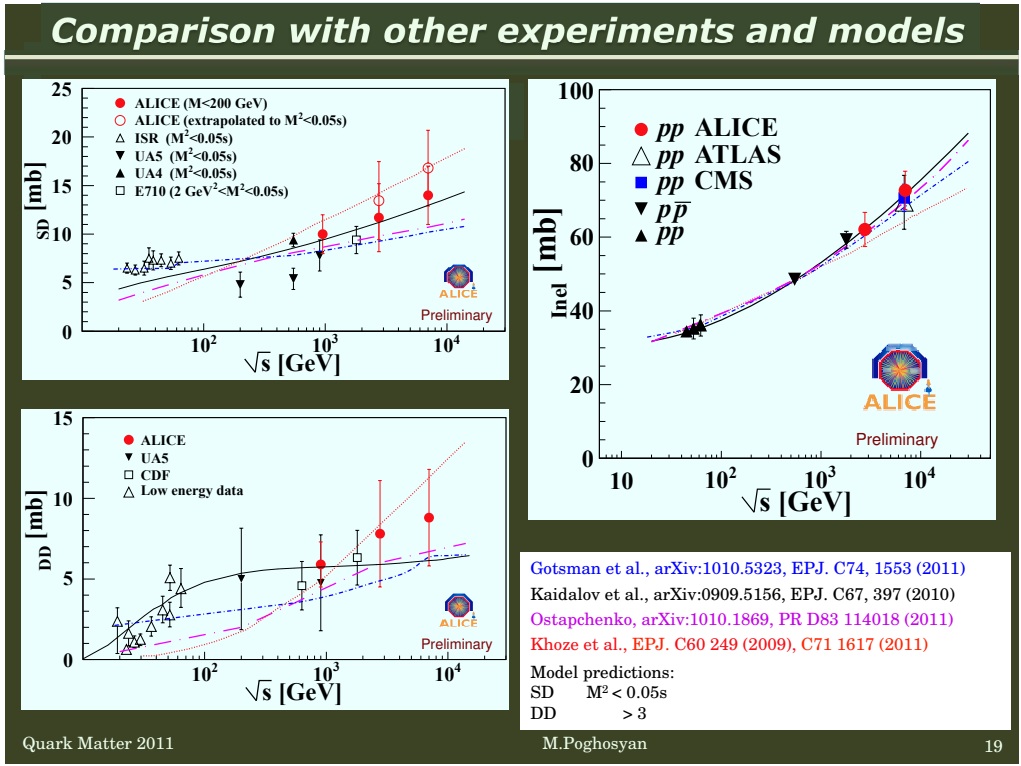}
\caption{Comparison of experiments and models - from M. Poghosyan, Quark
Matter 2011.} \label{compar_fig}
\end{figure*}

\section{Single Inclusive cross sections}
The first results emanating from the LHC was a measurement of the single 
inclusive cross section \cite{cms1},\cite{Atlas1},\cite{Alice1} .
As shown in Fig.4. by the CMS collaboration \cite{cms1}, the inclusive 
data appear to have a different energy dependence at lower ISR energies,
where it can be parametrized by a ln s behaviour, than at the higher LHC
energies where it has a ln$^{2}$s dependence.
This variance in energy dependence also appears in the results of Monte 
Carlo programs which gave a good description of the data when tuned for 
energies of
$W \leq 1800$ GeV and failed when extended to LHC energies \cite{field}.
This is clearly illustrated in Fig.5, which was taken from a talk given by 
Rick 
Field in Zakopane \cite{field}.
\begin{figure*} [ht]
\centering
\includegraphics [width=135mm]{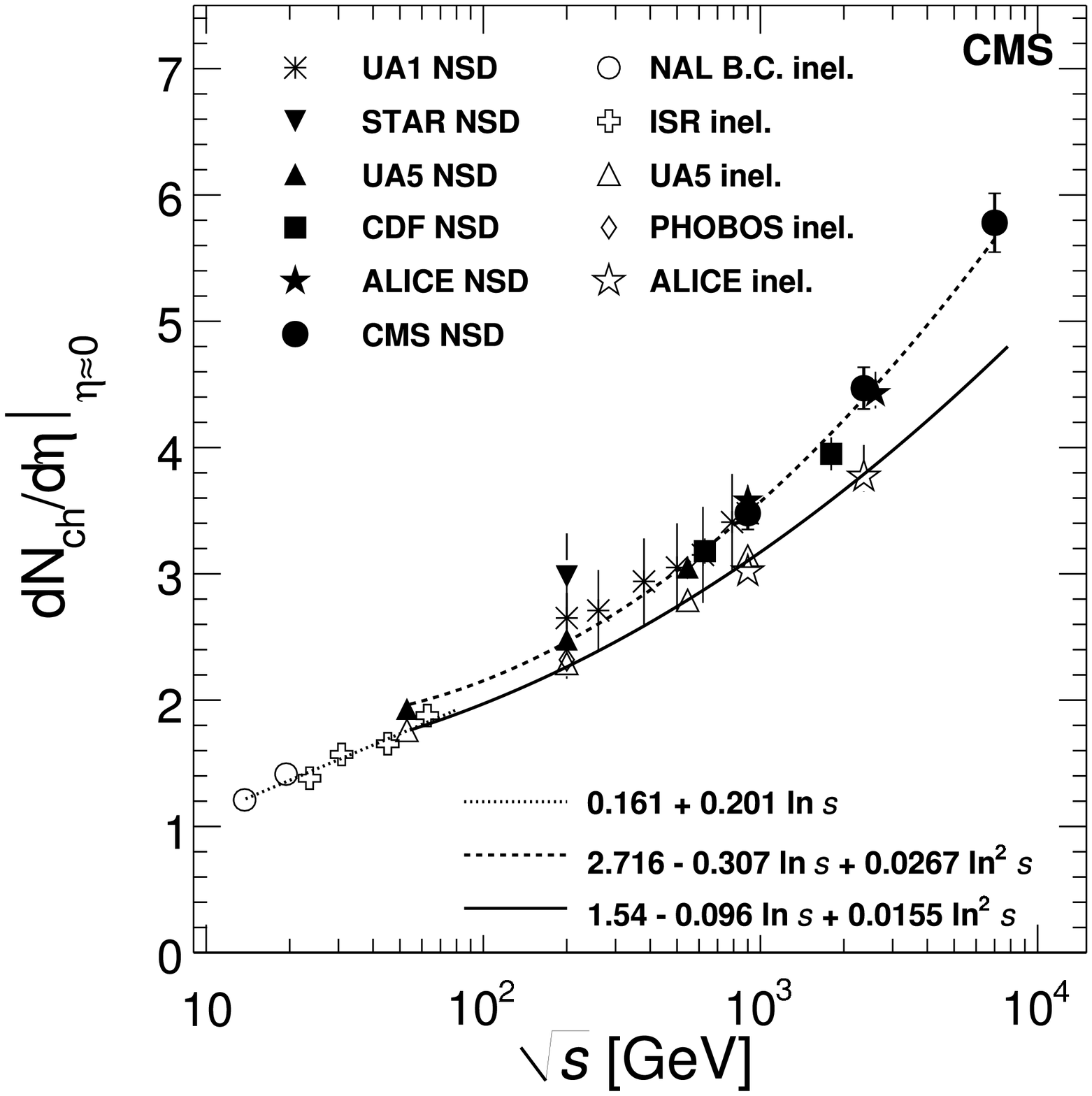}
\caption{Average value of $dN_{ch}/d \eta$ in the central $\eta$ region
from \cite{cms1}.} \label {figure4}
\end{figure*}

\begin{figure*} [ht]
\centering
\includegraphics [width=135mm]{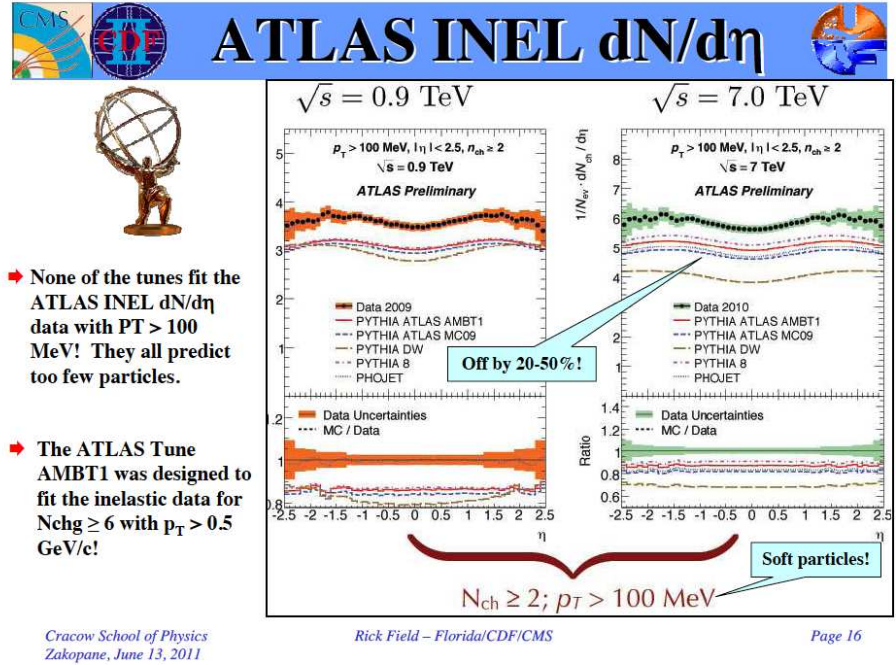}
\caption{Results of Monte Carlo tunes at $\sqrt{s}$ = 0.9 and 7.0 TeV 
from 
\cite{field}. } \label {figure5}
\end{figure*}

 We were curious to see if our model was compatible with this feature.
To test this we  extended our model of soft interactions so as to examine 
its reliability at LHC energies.

 The single inclusive cross section:
\begin{eqnarray}
&&\frac{1}{\sigma_{NSD}}\,\frac{d \sigma}{d y}\,\,
=\,\,\frac{1}{\sigma_{NSD}(Y)}\,\left\{ a_{\pom \pom}   
\Big( \int d^2 b \Big( \alpha^2 \,G_1(b, Y/2 - y)
+ \beta^2 G_2(b,Y/2 -y)\Big)\right. \nonumber\\
&&\left.\times\,\int d^2 b \Big( \alpha^2 \,G_1(b, Y/2 + y)
+ \beta^2 G_2(b,Y/2 +y)\Big)  \right. \nonumber\\
&&\left.\,\,\,-\,\,a_{ \pom \reg} \,\,( \alpha^2 \,g^\reg_1
+ \beta^2 g^\reg_2)\,\Big[\alpha^2 \,
\int d^2 b \Big( \alpha^2 \,G_1(b, Y/2 - y)
+ \beta^2 G_2(b,Y/2 -y)\Big)\,e^{\Delta_\reg\,(Y/2 + y)}\right.\nonumber\\
&&\left.+\,\int d^2 b \Big( \alpha^2 \,G_1(b, Y/2 + y)
+ \beta^2 G_2(b,Y/2 +y)\Big)\,e^{\Delta_\reg\,(Y/2 - y)}
\Big]\right\}.
\label{eq-sics}
\end{eqnarray}

\subsection{Comparison with LHC single inclusive data}
 To compare our model with data, we need to introduce two new 
phenomenological parameters $a_{\p \p}$ and $a_{R \p}$ which describe the 
emission of hadrons from the Pomeron and Reggeon, as well as the ratio
$\frac{Q_{0}}{Q}$ of the  
dimensional parameters $Q$ and $Q_{0}$. $Q$ is the average 
transverse momentum of the produced minijets, and $\frac{Q_{0}}{2}$ 
denotes the mass of the slowest hadron produced in the decay of the 
minijet. The three additional parameters are determined by fitting to 
experimental inclusive data.

\par We made two separate fits: \\
a) fitting to the CMS data at different LHC energies, and\\
b) fitting to all inclusive data for W $\geq$ 546 GeV. \\
The results of our fit are shown in Table 1.

\begin{table*} [t]
\begin{center}
\caption { Coefficients and results of fit to inclusive cross sections}
\begin{tabular}{|c|c|c|c|c|} \hline
Data & $a_{\pom \pom} $ &   $a_{\pom \reg}$  & $Q_{0}/Q$
 & $\chi^{2}/d.f.$ \\
\hline \hline
All  & 0.396  & 0.186 & 0.427 & 0.9\\ \hline
CMS  & 0.413 & 0.194 & 0.356 & 0.2\\ \hline \hline
\end{tabular}
\label{table_incldata}
\end{center}
\end{table*}

In Fig.6. we compare the predictions of our model (using the parameters 
given in Table 1) with the experimental results given  by the three LHC
experimental groups \cite{cms1},\cite{Atlas1},\cite{Alice1} and from PDG 
\cite{PDG}.

\begin{figure*}[ht]
\centering
\begin{tabular}{c c c}
\includegraphics[width =70mm]{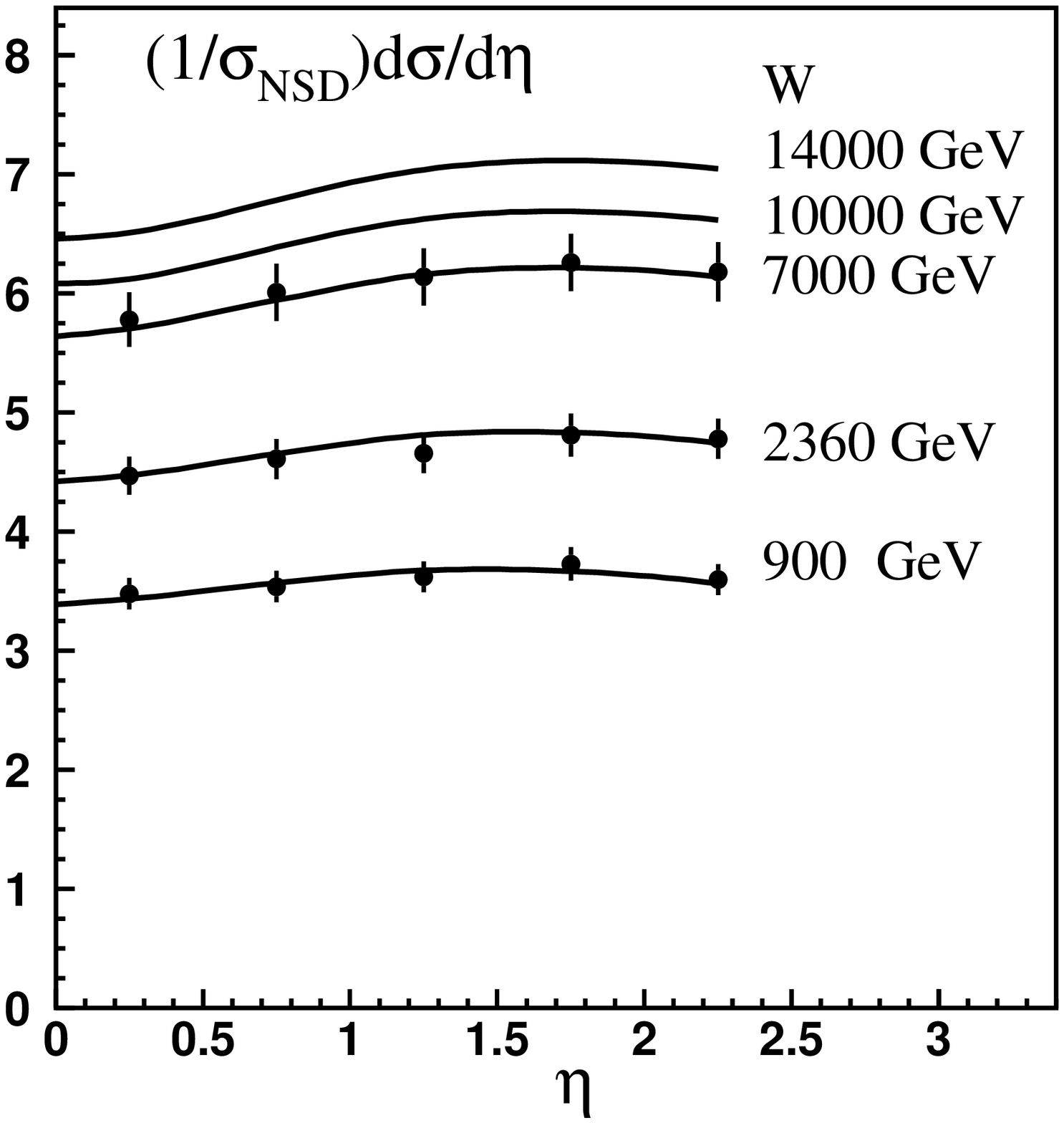} &\,\,\,\,&
\includegraphics[width =70mm]{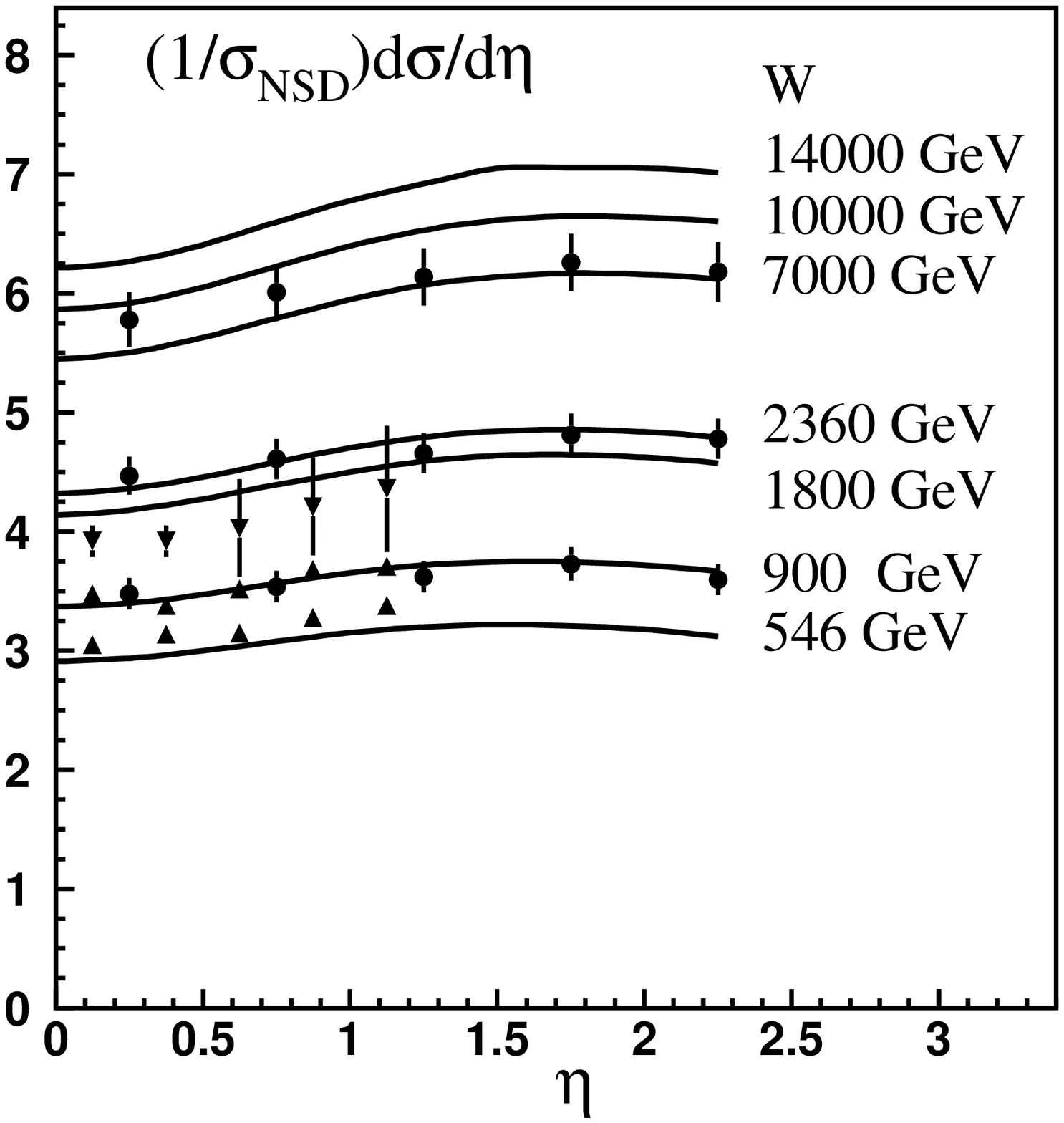}
\end{tabular}
\caption{Results of GLM for single inclusive cross section}
\label{incl_fig}
\end{figure*}
 
In Fig.6(a) (the left hand figure) we compare with the CMS data 
\cite{cms1},
while in Fig.6(b), (the right hand figure) we compare with all 
data available at energies $W \geq$ 546 GeV \cite{cms1},\cite{Atlas1}  
\cite{Alice1}, \cite{PDG}. 
The fit to the CMS data for energies 900 GeV $\leq W \leq$  7000 GeV is 
far superior to that made to all data 546 GeV $\leq W \leq$  7000 GeV. 
Whether this is due to normaliztion problems, or unascertained 
physical phenomena, is still an open question.  

\section{Summary and Conclusions}

\begin{itemize}
\item
We present a model for soft interactions having two components: \\ 
(i) G-W mechanism for elastic and low mass diffractive scattering. \\
(ii) Pomeron enhanced contributions for high mass diffractive production.
\item
 Enhanced $\pom$ diagrams, make important contributions
to both $\sigma_{sd}$ and $\sigma_{dd}$.
\item
Most phenomenological models which successfully describe LHC data, are
found lacking at lower energies.
This is shown explicitly in Fig.3, where the fit denoted Khoze \cite{KMR}
 which goes through
the LHC points for $\sigma_{sd}$ and $\sigma_{dd}$, completely underestimates
 the lower energy measurements.

\item
Monte Carlo generators which were successful in describing data for
W $\leq$ 1.8 TeV, need to be retuned to describe LHC data.
 This is illustrated in Fig.5.
\item
GLM model (with parameters determined by data for W $\leq$ 1.8 TeV )
underestimates inclusive rapidity distribution ($\sigma_{NSD}$)
data for   $\sqrt{s}$ = 7 TeV.\\
Fit made to LHC data (ONLY), is successful.
\end{itemize}

\par We trust that the enigma regarding the cause of the apparent change in
 energy behaviour at  
LHC energies, will be solved in the near future.

\begin{acknowledgments}
 I would like to thank my friends and collegues Genya Levin and Uri Maor, 
for an enjoyable  
 collaboration, the fruits of which are presented here.
\end{acknowledgments}

\bigskip 

\end{document}